\documentclass[showpacs,amsmath,amssymb,aps,showkeys,floatfix,prd,a4paper]{revtex4}

\usepackage[dvips]{graphicx}
\usepackage{dcolumn}
\usepackage{bm}
\usepackage{epsfig}
\usepackage{amsfonts}
\usepackage{amssymb,amscd}

\def\lsim{\raise0.3ex\hbox{$<$\kern-0.75em\raise-1.1ex\hbox{$\sim$}}}

\def\gsim{\raise0.3ex\hbox{$>$\kern-0.75em\raise-1.1ex\hbox{$\sim$}}}

\newcommand{\be}{\begin{equation}}

\newcommand{\ee}{\end{equation}}

\def\beq{\begin{equation}}

\def\eeq{\end{equation}}

\def\beqa{\begin{eqnarray}}

\def\eeqa{\end{eqnarray}}

\newcommand{\ba}{\begin{eqnarray}}

\newcommand{\ea}{\end{eqnarray}}

\def\gappeq{\mathrel{\rlap {\raise.5ex\hbox{$>$}}

{\lower.5ex\hbox{$\sim$}}}}

\def\lappeq{\mathrel{\rlap{\raise.5ex\hbox{$<$}}

{\lower.5ex\hbox{$\sim$}}}}

\def\Toprel#1\over#2{\mathrel{\mathop{#2}\limits^{#1}}}

\begin{document}

\title{Exclusive dilepton production in  ultraperipheral $PbPb$ collisions at the LHC}
\author{C. Azevedo$^1$, V.P. Gon\c{c}alves$^1$ and  B.D.  Moreira$^{1,2}$}
\affiliation{$^1$ High and Medium Energy Group, Instituto de F\'{\i}sica e Matem\'atica,  Universidade Federal de Pelotas (UFPel)\\
Caixa Postal 354,  96010-900, Pelotas, RS, Brazil.\\
$^2$ Departamento de F\'isica, Universidade do Estado de Santa Catarina, 89219-710 Joinville, SC, Brazil.  \\
}

\begin{abstract}
In this paper we perform a systematic study of the  exclusive dilepton production by $\gamma \gamma$ interactions in $PbPb$ collisions at the LHC Run 2 energies considering different levels of precision for the treatment of the absorptive corrections and for the nuclear form factor. The rapidity and invariant mass distributions are estimated taking into account the experimental cutoffs and a comparison with the recent ALICE and ATLAS data for the $e^+ e^-$ and $\mu^+ \mu^-$ production is presented.
\end{abstract}



\maketitle

\vspace{1cm}

%\section{Introduction}

\begin{figure}
%\centerline{\psfig{figure=gamavec.eps,width=10cm}}  
\centerline{\psfig{figure=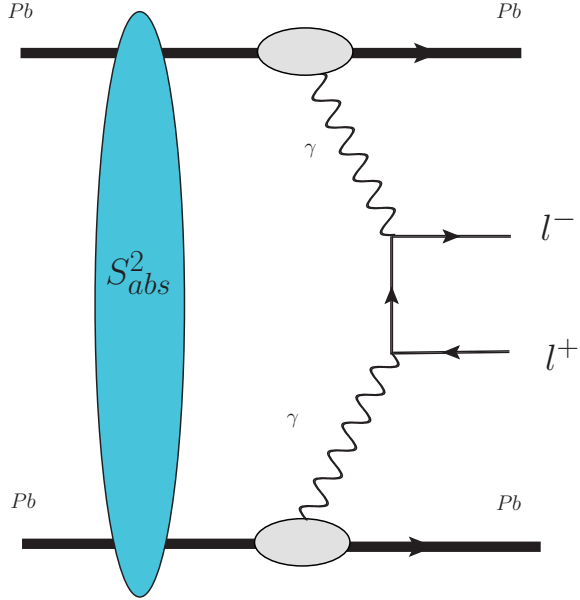,width=8cm}}
\caption{Exclusive dilepton production by $\gamma \gamma$ interactions in ultraperipheral $PbPb$ collisions.}
\label{fig:diagram}
\end{figure}

The recent experimental results from RHIC \cite{star_dilepton} and 
LHC \cite{alice_dilepton,atlas_dilepton, cms_dilepton}  demonstrated that the study of the dilepton production by photon -- photon  interactions in ultraperipheral heavy ion collisions (UPHICs) is  feasible. Such collisions are characterized by an impact parameter $b$ greater than the sum of the radius of the incident nuclei, which implies the suppression of the strong interactions and the dominance of the electromagnetic interaction between them \cite{upc,upc_dilepton}. The intense electromagnetic fields that accompany the relativistic heavy ions can be viewed as a spectrum of equivalent photons and the dileptons can be produced through  the $\gamma \gamma \rightarrow l^+ l^-$ process (See Fig. \ref{fig:diagram}). one has that the photon flux is proportional to the square of the nuclear charge $Z$ and the associated cross section  to $Z^4$, implying large  cross sections at RHIC and LHC energies. Moreover, the final state is very simple, consisting  of a dilepton pair with very small transverse momentum and  two intact nuclei. Such processes are usually denoted exclusive and are characterized by two rapidity gaps, i.e. empty regions  in pseudo-rapidity that separate the intact very forward nuclei from the $l^+ l^-$ state. Such aspects have motivated the experimental analyzes performed 
by the STAR \cite{star_dilepton}, ALICE \cite{alice_dilepton}, ATLAS \cite{atlas_dilepton} and CMS \cite{cms_dilepton} Collaborations as well as the 
 improvement of the theoretical description of this process.

 Two important aspects in the treatment of the exclusive dilepton production  are the description of the absorptive effects, that suppress the strong interactions, and the modelling of the nuclear form factor that determines the equivalent photon flux of the nuclei. In general, the suppression is performed by expressing the cross section in the impact parameter representation and including an absorptive factor $S^2_{abs}(b)$ in its calculation. Such factor is dependent on the impact parameter and, for the case of a collision between identical nuclei,  it is equal to zero for $b \ll 2R$ and equal to 1 for $b \gg 2R$, where $R$ is the nuclear radius. However, its treatment in the region where $b \approx 2R$ is still an open question, and different authors assume distinct levels of precision in the calculation of these quantities 
  \cite{Baur_Ferreira, Cahn, Baur_Baron, Baltz_Klein,super}. 
On the other hand, the equivalent photon flux and its dependence on the impact parameter is determined by the modelling of the  form factor $F(q)$, which is the Fourier transform of the charge distribution of the nucleus. As  demonstrated in Refs. \cite{greiner,mariola},  the behavior of the equivalent photon flux for $b \lesssim R$ is strongly  dependent on the model assumed for $F(q)$. As the maximum value of the photon energy $\omega$, associated to the electromagnetic field of the relativistic ion, is proportional to the Lorentz gamma factor $\gamma_L$ and inversely proportional to the impact parameter, one has that the treatment of flux at small $b$ has direct impact on the estimates of the production of dileptons with large invariant mass $W$. The basic  motivation of the study performed in this paper is to estimate the impact of distinct levels of precision in the calculation of  $S^2_{abs}(b)$ and $F(q)$ on the predictions of the rapidity and invariant mass dilepton  distributions. In particular, we will take into account the ALICE and ATLAS experimental cuts, and will estimate the exclusive $e^+ e^-$ and $\mu^+ \mu^-$ production by $\gamma \gamma$ interaction in $PbPb$ collisions at the LHC energies. Finally, a comparison with the recent data will be presented.

%The content of this paper is organized as follows. In the next section we present a brief review of the formalism for the treatment of the dilepton production by photon - photon   interactions in ultraperipheral heavy ion collisions. In Section \ref{results} we present our predictions for the rapidity and  invariant mass distributions considering the different models for the absorptive factor and nuclear form factor. Moreover, a comparison of these predictions with the recent ALICE and ATLAS experimental data is also presented. Finally, in Section \ref{conc} we summarize our main conclusions.

%\section{Formalism}
Initially, let's present a brief review of the main formulas to describe the exclusive dilepton production by $\gamma \gamma$ interactions in ultraperipheral $PbPb$ collisions, represented in Fig. \ref{fig:diagram}. Considering the Equivalent Photon Approximation (EPA) \cite{upc,epa} and assuming the impact parameter representation, one has that the total cross section for this process can be written as
\begin{eqnarray}
\sigma \left( Pb Pb \rightarrow Pb \otimes l^+ l^- \otimes Pb ;s \right)   
&=& \int \mbox{d}^{2} {\mathbf b_{1}}
\mbox{d}^{2} {\mathbf b_{2}} 
\mbox{d} \omega_{1}
\mbox{d} \omega_{2} \,\, \hat{\sigma}\left(\gamma \gamma \rightarrow l^+ l^- ; 
W \right )  N\left(\omega_{1},{\mathbf b_{1}}  \right )
 N\left(\omega_{2},{\mathbf b_{2}}  \right ) S^2_{abs}({\mathbf b})  
  \,\,\, ,
\label{cross-sec-1}
\end{eqnarray}
where $\sqrt{s}$ is center - of - mass energy of the $PbPb$ collision, $\otimes$ characterizes a rapidity gap in the final state and 
$W = \sqrt{4 \omega_1 \omega_2}$ is the invariant mass of the $\gamma \gamma$ system. The cross section $\hat{\sigma}$ is the elementary cross section to produce a pair of leptons with mass $m_l$, which can be calculated using the Breit - Wheller formula. Moreover, $N(\omega_i, {\mathbf b}_i)$ is the equivalent photon spectrum  
of photons with energy $\omega_i$ at a transverse distance ${\mathbf b}_i$  from the center of nucleus, defined in the plane transverse to the trajectory. The spectrum can be 
expressed in terms of the charge form factor $F(q)$ as follows \cite{upc}
\begin{eqnarray}
 N(\omega_i,b_i) = \frac{Z^{2}\alpha_{em}}{\pi^2}\frac{1}{b_i^{2} v^{2}\omega_i}
\cdot \left[
\int u^{2} J_{1}(u) 
F\left(
 \sqrt{\frac{\left( \frac{b_i\omega_i}{\gamma_L}\right)^{2} + u^{2}}{b_i^{2}}}
 \right )
\frac{1}{\left(\frac{b_i\omega_i}{\gamma_L}\right)^{2} + u^{2}} \mbox{d}u
\right]^{2} \,\,.
\label{fluxo}
\end{eqnarray}
 The factor $S^2_{abs}({\mathbf b})$ depends on the impact parameter ${\mathbf b}$ of the $PbPb$ collision and  is denoted the absorptive  factor, which excludes the overlap between the colliding nuclei and allows to take into account only ultraperipheral collisions.
 Remembering that the photon energies $\omega_1$ and $\omega_2$  are related to   
$W$ and to the rapidity  $Y = \frac{1}{2}(y_{l^+} + y_{l^-})$ of the outgoing dilepton pair system by 
\begin{eqnarray}
\omega_1 = \frac{W}{2} e^Y \,\,\,\,\mbox{and}\,\,\,\,\omega_2 = \frac{W}{2} e^{-Y} \,\,\,,
\label{ome}
\end{eqnarray}
one has that 
the total cross section can be expressed by (For details see e.g. Ref. \cite{mariola})
\begin{eqnarray}
\sigma \left(Pb Pb \rightarrow Pb \otimes l^+ l^- \otimes Pb;s \right)   
&=& \int \mbox{d}^{2} {\mathbf b_{1}}
\mbox{d}^{2} {\mathbf b_{2}} 
\mbox{d}W 
\mbox{d}Y \frac{W}{2} \, \hat{\sigma}\left(\gamma \gamma \rightarrow l^+ l^- ; 
W \right )  N\left(\omega_{1},{\mathbf b_{1}}  \right )
 N\left(\omega_{2},{\mathbf b_{2}}  \right ) S^2_{abs}({\mathbf b})  
  \,\,\, .
\label{cross-sec-2}
\end{eqnarray}
It is important to emphasize that in EPA we disregard the photon virtualities, which is a good approximation, mainly for ions, since the typical virtualities are 
$< 1/R$. Moreover, as already emphasized,  the highest energy of the photons is of the order of the inverse Lorentz contracted radius of the nuclei 
$\approx \gamma_L/R$, with the spectra decreasing exponentially at larger energies.

\begin{figure}
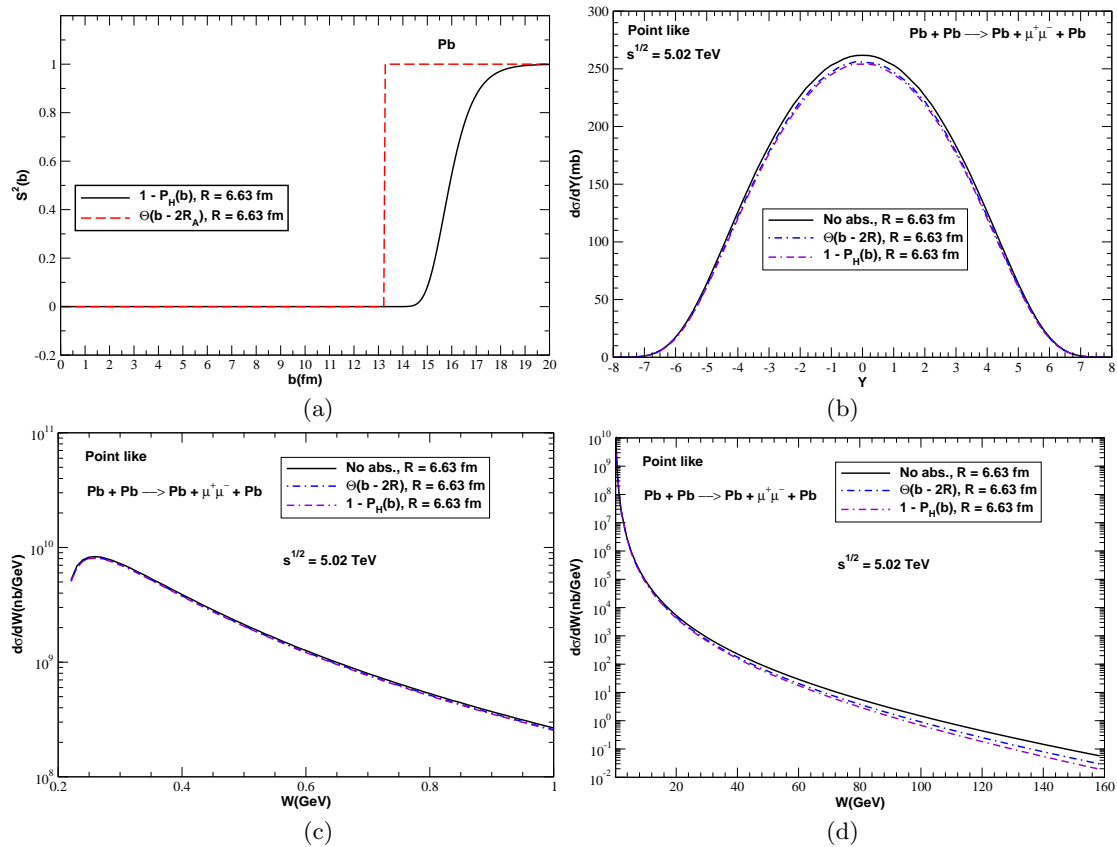

\begin{tabular}{cc}
\hspace{-1cm}
{\psfig{figure=Pb_probabilidade_v2.eps,width=7.2cm}} & 
{\psfig{figure=distribuicao_de_rapidez_point_mb_v2.eps,width=7.2cm}} \\
(a) & (b) \\ 
\hspace{-1cm}
{\psfig{figure=dsigdw-full-point-LM-v2.eps,width=7.2cm}} & {\psfig{figure=dsigdw-full-point-HM-v2.eps,width=7.2cm}} \\
(c) & (d) %{\psfig{figure=pp-psi-psi-500.eps,width=5cm}} & 
%{\psfig{figure=pp-psi-psi-7000.eps,width=5cm}} & 
%{\psfig{figure=pp-psi-psi-14000.eps,width=5cm}}
\end{tabular}                                                                                                                       
\caption{(a) Impact parameter dependence of the absorptive factor; (b) Rapidity distribution for the exclusive dimuon production by $\gamma \gamma$ interactions in $PbPb$ collisions at $\sqrt{s} = 5.02$ TeV considering the different models for the absorptive factor $S^2_{abs}$; (c) Invariant mass distribution of the dimuons predicted by the different models of $S^2_{abs}$; (d) Idem for a larger invariant mass range.}
\label{fig:absoptive}
\end{figure}

The calculation of the invariant mass and rapidity distributions for the dilepton production in ultraperipheral heavy ion collisions requires as input  the nuclear form factor $F$, which determines the equivalent photon flux, and the absorptive factor $S^2_{abs}$.  The Eq. (\ref{cross-sec-2}) have been used by several authors in literature considering distinct levels of precision in the calculation of $F$ and $S^2_{abs}$
\cite{Baur_Baron,Baltz_Klein,mariola,mariola2,mariola3,zha1,zha2,guclu1,guclu2,
klein_peripheral}. Moreover, different values of the lower integration limits in ${\mathbf b_{1}}$ and ${\mathbf b_{2}}$ are sometimes assumed in the calculations, as well distinct values for the nuclear radius. In particular, some authors (See e.g. \cite{klein_peripheral}) assume the requirement that $b_i > R$, which ensure that the final state is produced outside of the nuclei and, consequently, does not interact with the nucleus. For hadronic final states,  interactions inside the nucleus can suppress the cross section and break it, which justifies this approximation. However, for dileptons, this requirement is not necessary. Therefore, it is important to estimate the impact of this approximation.   The presence of different assumptions in the distinct calculations in the literature, implies that a direct comparison between its predictions and  the interpretation of the results are not, in general, an easy task. In what follows we will estimate the impact of these distinct assumptions on the invariant mass and rapidity distributions. Initially, let's consider the modelling of the absorptive factor
$S^2_{abs}$. Baur and Ferreira - Filho proposed in Ref. \cite{Baur_Ferreira} to exclude the 
strong interactions between the incident nuclei by assuming that 
\begin{eqnarray}
S^2_{abs}({\mathbf b}) = \Theta\left(
\left|{\mathbf b}\right| - 2 R
 \right )  = 
\Theta\left(
\left|{\mathbf b_{1}} - {\mathbf b_{2}}  \right| - 2 R
 \right )  \,\,,
\label{abs1}
\end{eqnarray}
where $R$ is the nuclear radius. Such equation treats the nuclei as hard spheres with radius $R$ and assumes that the probability to have a hadronic interaction when $b > 2 R$ is zero. A more realistic treatment, which takes into account that  this probability is finite at  $b \gtrsim 2 R$, can be obtained using the Glauber formalism. In this case, 
$S^2_{abs}({\mathbf b})$ can be expressed in terms of the probability of interaction between the nuclei at a given impact parameter, $P_{H}({\mathbf b})$, being given by \cite{Baltz_Klein}
\begin{eqnarray}
 S_{abs}^{2} ({\mathbf b}) 
 =
 1 - P_{H}({\mathbf b}) 
\label{abs2}
 \end{eqnarray}
where
\begin{eqnarray}
 P_{H}({\mathbf b}) = 1 - \exp\left[
 - \sigma_{nn} \int d^{2} {\mathbf r} 
 T_{A}({\mathbf r}) T_{A}({\mathbf r} - {\mathbf b})
 \right].
\end{eqnarray}
with $\sigma_{nn}$ being the total hadronic interaction cross section and $T_{A}$ the nuclear thickness function. As in Ref. \cite{Baltz_Klein} we will assume that $\sigma_{nn} = 88$ mb at the LHC. In Fig. \ref{fig:absoptive} (a) we present a comparison between the predictions for 
$ S_{abs}^{2}$ from Eqs. (\ref{abs1}) and (\ref{abs2}). In particular, we show the results from Eq. (\ref{abs1}) for a nuclear radius given by $R = 6.63$ fm, which is value determined in low - energy electron scattering experiments. We have that the main difference is that the description of the absorptive factor given by Eq. (\ref{abs2}) implies a smooth transition between the small ($b \ll 2R$)  and large ($b \gg 2R$) impact parameter behaviours. The impact of these distinct approaches on the rapidity and invariant mass distributions for the exclusive dimuon production in $PbPb$ collisions at $\sqrt{s} = 5.02$ TeV is presented in Figs. \ref{fig:absoptive} (b), (c) and (d). In this calculation we have assumed a point - like form factor ($F = 1$) and $b_i^{min} = R$ ($i = 1,2$). Such approximations are present in the STARlight Monte Carlo \cite{starlight}, which is usually considered in the analysis of the exclusive dilepton production in $AA$ collisions. 
Moreover,  kinematical cutoffs in $Y$ and $W$ were not included. For completeness, the prediction obtained disregarding the absorptive effects ($ S_{abs}^{2} = 1$) is also presented. The results for the rapidity distribution are shown in Fig. \ref{fig:absoptive} (b) indicate that the treatment of the absorptive corrections modify the magnitude of the distribution at midrapidities ($Y \approx 0$). As expected, the inclusion of the absorptive factor reduces the cross section. On the other hand, the predictions obtained using Eqs. (\ref{abs1}) and (\ref{abs2}) are similar. In Figs. \ref{fig:absoptive} (c) and (d) we present our predictions for the invariant mass distribution considering the small and large invariant mass ranges, respectively. We have that at small invariant masses, the predictions of the different approaches are similar. In contrast, its predictions are distinct at large $W$. Such results are expected, since the main contribution for the production of dimuons with small invariant mass comes from photons with small energy and large $b_i$ (See e.g. \cite{guclu2}), where the predictions of the different approaches for the absorptive factor are almost identical. On the other hand, the production of dileptons with large invariant mass is associated to photons with large energy and $b_i \approx R$, where the treatment of $ S_{abs}^{2}$ is more model dependent. We have that the $ S_{abs}^{2}$ given by Eq. (\ref{abs2}), which is a more realistic approach, predicts smaller values of the distribution for large $W$.

\begin{figure}
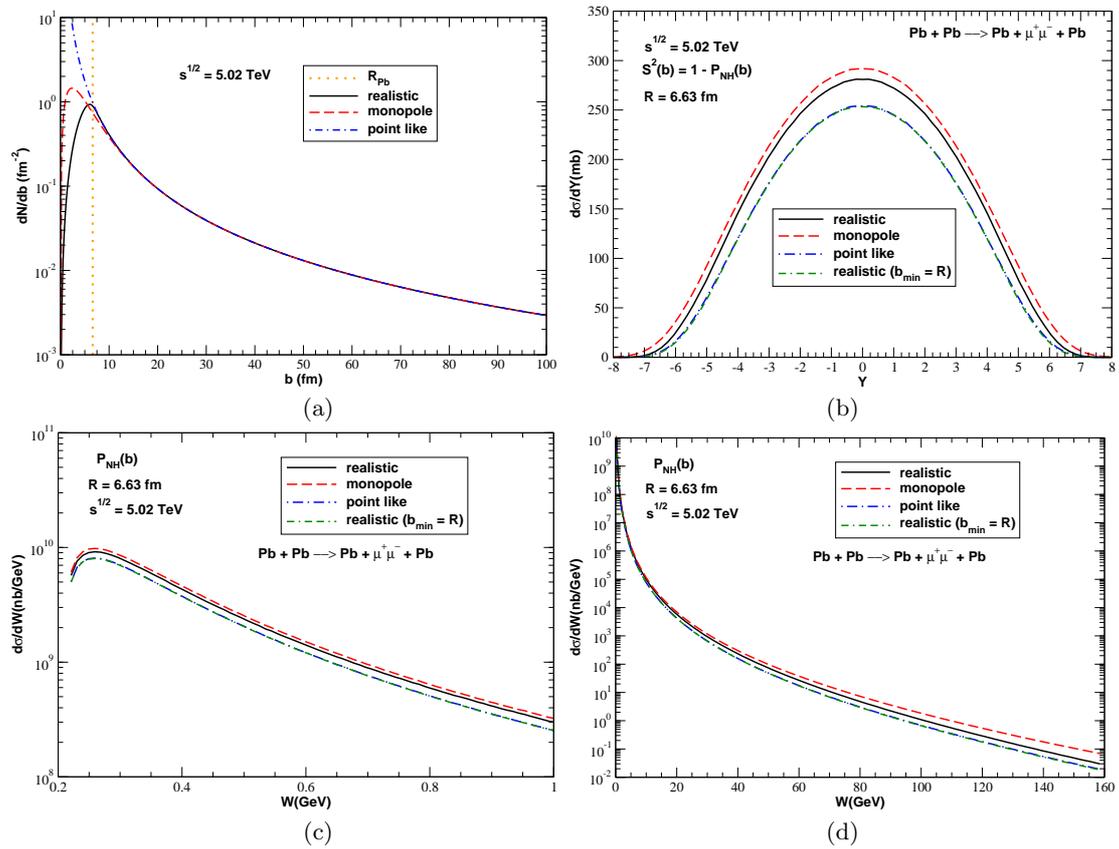

\begin{tabular}{cc}
\hspace{-1cm}
{\psfig{figure=fluxo_NEW.eps,width=7.2cm}} & 
{\psfig{figure=distribuicao_de_rapidez_S2_PNH_mb_TEST_bmin.eps,width=7.2cm}} \\ 
(a) & (b) \\ 
\hspace{-1cm}
{\psfig{figure=dsigdw-full-S2_PNH_bmin_2.eps,width=7.2cm}}&
{\psfig{figure=dsigdw-full-S2_PNH_bmin.eps,width=7.2cm}} \\
(c) & (d) \\ 
%{\psfig{figure=pp-psi-psi-500.eps,width=5cm}} & 
%{\psfig{figure=pp-psi-psi-7000.eps,width=5cm}} & 
%{\psfig{figure=pp-psi-psi-14000.eps,width=5cm}}
\end{tabular}                                                                                                                       
\caption{(a) Impact parameter dependence of the equivalent photon flux; (b) Rapidity distribution for the exclusive dimuon production by $\gamma \gamma$ interactions in $PbPb$ collisions at $\sqrt{s} = 5.02$ TeV considering the different models for nuclear form factor  $F(q)$; (c) Invariant mass distribution of the dimuons predicted by the different models of $F(q)$; (d) Idem for a large invariant mass range.}
\label{fig:fluxo}
\end{figure}

Let's now analyze the impact of the different levels of precision in the treatment  of the nuclear form factor on the rapidity and invariant mass distributions. As discussed before, the modelling of $F$ determines the impact parameter dependence of the equivalent photon flux. In our study we will consider the point - like  ($F = 1$) and  the monopole form factor, which is given by $F(q^2) = \Lambda^2/(\Lambda^2 + q^2)$, with $\Lambda = 0.088$ GeV adjusted to reproduce the root - mean - square (rms) radius of the nucleus \cite{mariola}.
In addition we will consider the realistic form factor, which corresponds to the Wood - Saxon distribution and is the Fourier transform of the charge density of the nucleus, being analytically expressed by
\begin{eqnarray}
 F(q^{2}) = 
 \frac{4\pi\rho_{0}}{Aq^{3}} 
 \left[ 
 \sin(qR) - qR \cos(qR) 
 \right]
 \left[
 \frac{1}{1 + q^{2} a^{2}}
 \right]
\end{eqnarray}
with $a = 0.549$ fm and $R_{A} = 6.63$ fm \cite{DeJager:1974liz,Bertulani:2001zk}. As discussed in detail in Ref. \cite{mariola}, the monopole and realistic form factors are similar  in a limited range of $q$ and are  distinct at large $q$. 
It is important to emphasize that the point - like form factor is an unrealistic approximation, since disregards the internal structure of the nucleus. The monopole is more precise, but still is an approximation for the realistic approach derived using the Wood - Saxon distribution, which corresponds to a charge density constrained by the experimental data.
In Fig. \ref{fig:fluxo} (a) we present the predictions for the impact parameter dependence of the corresponding equivalent photon fluxes.  We have that the predictions are similar at large $b$ but differ at small values of the impact parameter. While the point - like prediction is singular for $b \rightarrow 0$, the monopole and realistic fluxes are finite. As a consequence, for these two models we can assume the lower limits of $b_i$ integrations present in Eq. (\ref{cross-sec-2}) as being zero. In the point - like we will assume that $b_i^{min} = R$. For comparison, we also will present our results for the rapidity and invariant mass distributions obtained using the  realistic form factor and $b_i^{min} = R$. Our predictions for the exclusive dimuon production in $PbPb$ collisions at $\sqrt{s} = 5.02$ TeV are presented in Fig. \ref{fig:fluxo} (b), (c) and (d) considering that the absorptive factor is given by Eq. (\ref{abs2}) and $R = 6.63$ fm. The distinct predictions for the rapidity distribution are shown in Fig. \ref{fig:fluxo} (b).  We have that the behaviour of the distribution at midrapidities is strongly dependent on the approach used for the nuclear form factor, with the monopole one predicting the higher value. The realistic prediction is 3\% smaller than the monopole one.  One other hand, if in addition of the realistic form factor we also assume $b_i^{min} = R$, the prediction is suppressed by 17 \%. Such result demonstrate the importance of a correct treatment of the form factor in the region $b_i \lesssim R$ and that the assumption $b_i > R$ in the calculation of the  dilepton production  is not a good approximation.  The corresponding predictions for the invariant mass distributions at small and large invariant masses are presented in Figs. \ref{fig:fluxo} (c) and (d), respectively. We have that the predictions are similar for $W \le 1$ GeV and differ significantly at large $W$. As discussed before, such behaviour is expected, since the production of dimuons with a large invariant mass is generated by energetic photons present at small $b_i$, where the equivalent photon fluxes are distinct.

\begin{figure}
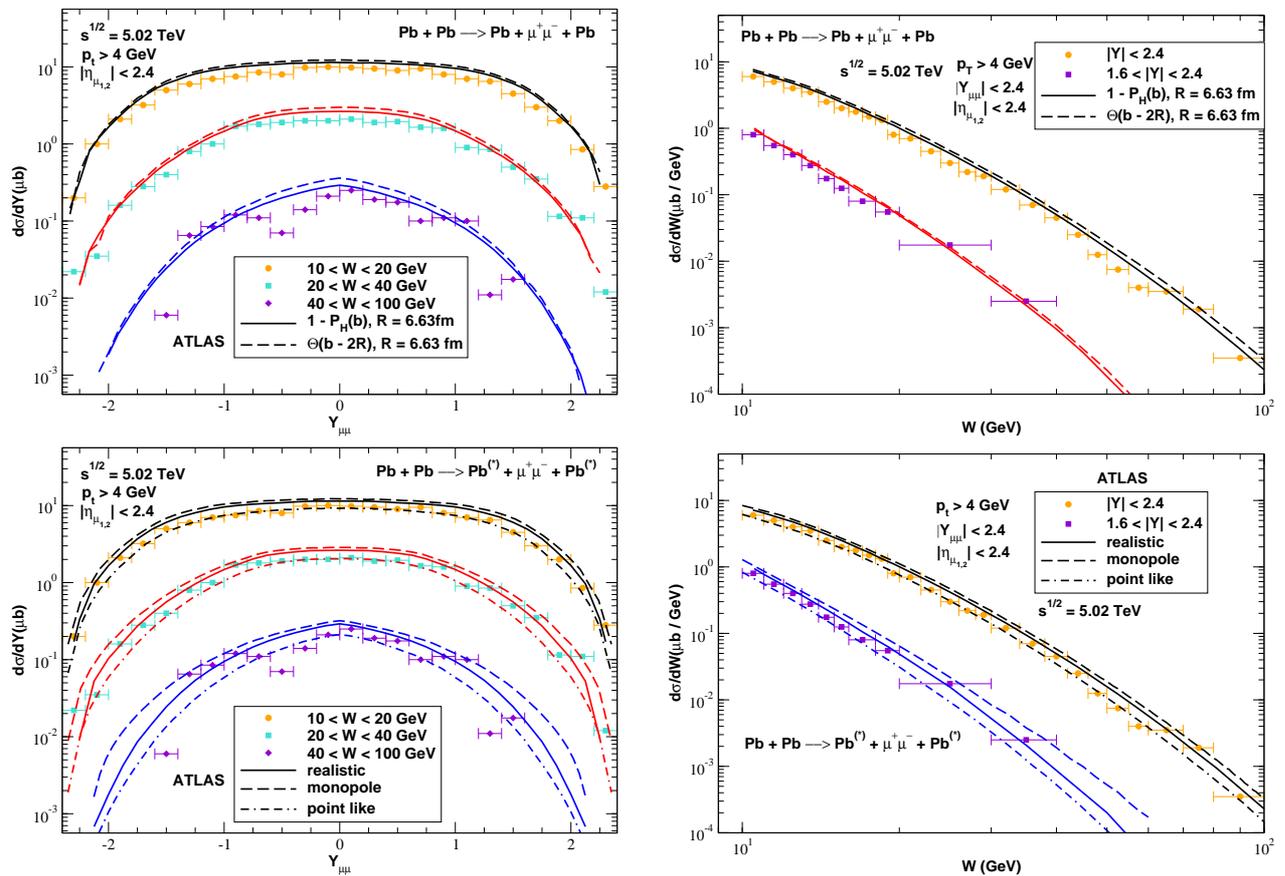

\begin{tabular}{cc}
{\psfig{figure=realistic_rapidez_dimuons_cortes_5020.eps,width=8cm}} & 
\hspace{0.5cm}{\psfig{figure=realistic_dist_W_muon.eps,width=8cm}}\\
{\psfig{figure=rapidez_dimuons_cortes_5020_NEW.eps,width=8cm}} & 
\hspace{0.5cm}{\psfig{figure=dist_W_muon_NEW.eps,width=8cm}}%& 
\end{tabular}                                                                                                                       
\caption{Rapidity and invariant mass distributions for the exclusive $\mu^{+}\mu^{-}$ production by $\gamma \gamma$ interactions in $PbPb$ collisions at $\sqrt{s} = 5.02$ TeV. Preliminary experimental data from the ATLAS Collaboration \cite{atlas_dilepton}. The predictions obtained considering different assumptions for the absorptive factor (nuclear form factor) are presented in the upper (lower) panels. }
\label{fig:atlas}
\end{figure}

\begin{figure}
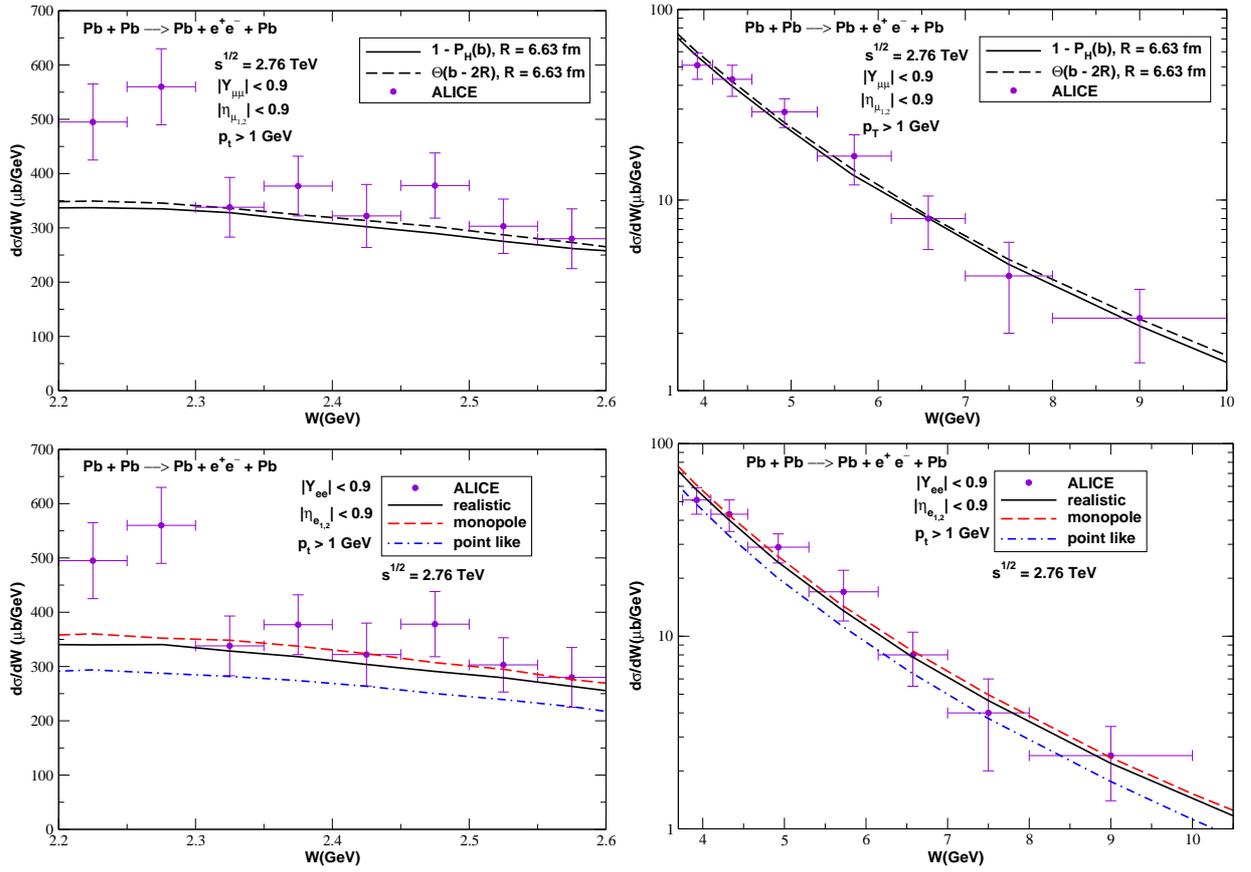

\begin{tabular}{cc}
{\psfig{figure=realistic_distW_2_2-2_6.eps,width=8cm}} & 
{\psfig{figure=realistic_distW_3_7-10.eps,width=8cm}} \\
{\psfig{figure=distW_2_2-2_6_NEW.eps,width=8cm}} & 
{\psfig{figure=distW_3_7-10_NEW.eps,width=8cm}} 
%{\psfig{figure=pA-psi-psi-5000.eps,width=7cm}} & 
%{\psfig{figure=pA-psi-psi-8800.eps,width=7cm}}
\end{tabular}                                                                                                                       
\caption{Rapidity and invariant mass distributions for the exclusive $e^{+}e^{-}$ production by $\gamma \gamma$ interactions in $PbPb$ collisions at $\sqrt{s} = 2.76$ TeV. Experimental data from the ALICE Collaboration \cite{alice_dilepton}. The predictions obtained considering different assumptions for the absorptive factor (nuclear form factor) are presented in the upper (lower) panels.}
\label{fig:alice}
\end{figure}

In the last years, the ALICE \cite{alice_dilepton} and ATLAS \cite{atlas_dilepton} Collaborations have release data for the dielectron and dimuon  production by $\gamma \gamma$ interactions in $PbPb$ collisions at $\sqrt{s} = 2.76$ and 5.02 TeV, respectively. In what follows we will extend our previous analyzes for the kinematical range probed by these experiments. Following Ref. \cite{russos} we will include the experimental cutoffs in rapidity, transverse momentum and invariant mass of the dilepton pairs and will compare the predictions, obtained considering different levels of precision in the calculation of the nuclear form factor and  the absorptive correction, with the experimental data. In Fig. \ref{fig:atlas} we compare our results with the ATLAS data for the dimuon production considering different models for the absorptive factor (upper panels) and for the nuclear form factor (lower panels). In our calculations we impose cuts on muon pseudo - rapidities, $-2.4 < \eta_{i,\mu} < 2.4$ and muon transverse momenta $p_t > 4.0$ GeV. 
It is important to emphasize that the ATLAS data include events where the ions dissociates in the collision.
Initially, let's estimate the impact of different models for $ S_{abs}^{2}$.    
The realistic form factor is assumed to calculate the invariant mass and rapidity distributions. We have that the data are well described by the EPA approach and that the distinct predictions are similar in the kinematical range covered by the ATLAS data. The difference between them increases when the invariant mass range covers larger values of $W$, which is expected from the analysis performed before without the inclusion of the kinematical cutoffs. Considering distinct assumptions for the nuclear form factor and the absorptive factor given by Eq. (\ref{abs2}), we have estimated the rapidity and invariant mass distributions, with the predictions being presented in the lower panels of Fig. \ref{fig:atlas}. We have that the predictions are dependent on the level of precision used to calculate the photon flux, with the monopole (point - like) predicting larger (smaller) values for the distributions in comparison to the more precise prediction derived using the realistic flux. The difference between the predictions increases at larger invariant masses. Similar conclusions can be derived from the analysis of the Fig. \ref{fig:alice}, where we compare our predictions with the experimental data from ALICE Collaboration for the dielectron production in $PbPb$ collisions at $\sqrt{s} = 2.76$ GeV. As in the experimental analysis, we impose  a cut on dielectron pair rapidity, $|Y_{ee}| < 0.9$, on the electron pseudo - rapidities, $-0.9 < \eta_{i,e} < 0.9$ and electron transverse momenta $p_t > 1.0$ GeV.    We have that the experimental data are satisfactorially described by the EPA approach, with exception of the two experimental points for the dielectron production with $W \le 2.3$ GeV. The analysis from Figs. \ref{fig:atlas} and \ref{fig:alice} indicate that the treatment of the absorptive factor has a small impact in the kinematical range probes by the ALICE and ATLAS experiments. On the other hand, the predictions at large invariant masses and large rapidities depend on the level of precision considered in the calculation of the nuclear form factor.

\vspace{1cm}

Finally, let us summarize our main conclusions. The recent experimental data from the STAR, ALICE, ATLAS and CMS Collaborations for the dilepton production  motivated a review of the main assumptions present in the theoretical approaches. 
In particular, in this paper we have estimated the impact of the different levels of precision used in the treatment of the absorptive effects and of the nuclear form factor on the predictions of the rapidity and invariant mass distributions. We have estimated these distributions without and with the inclusion of the experimental cutoffs and demonstrated that the distinct treatments for the absorptive factor have a small impact on the predictions. On the other hand, the correct treatment of the nuclear form factor is fundamental to obtain more precise predictions of the distributions at large invariant mass and large rapidities.

\begin{acknowledgments}
VPG acknowledge very useful discussions about photon - induced interactions with Spencer Klein, Mariola Klusek-Gawenda, Daniel Tapia - Takaki and Antoni Szczurek.
This work was  partially financed by the Brazilian funding
agencies CNPq, CAPES,  FAPERGS and INCT-FNA (process number 
464898/2014-5).
\end{acknowledgments}

\hspace{1.0cm}

\end{document}